# Estimation in hidden Markov models via efficient importance sampling

CHENG-DER FUH[1] and INCHI HU[2]

[1] *Graduate Institute of Statistics, National Central University, Chongli, Taiwan, Republic of China and Institute of Statistical Science, Academia Sinica, Nakang, Taipei 115, Taiwan, Republic of China. E-mail: stcheng@stat.sinica.edu.tw*

[2] *Department of Information and Systems Management, Hong Kong University of Science and Technology, Clear Water Bay, Kowloon, Honh Kong. E-mail: imichu@ust.hk*

Given a sequence of observations from a discrete-time, finite-state hidden Markov model, we would like to estimate the sampling distribution of a statistic. The bootstrap method is employed to approximate the confidence regions of a multi-dimensional parameter. We propose an importance sampling formula for efficient simulation in this context. Our approach consists of constructing a locally asymptotically normal (LAN) family of probability distributions around the default resampling rule and then minimizing the asymptotic variance within the LAN family. The solution of this minimization problem characterizes the asymptotically optimal resampling scheme, which is given by a tilting formula. The implementation of the tilting formula is facilitated by solving a Poisson equation. A few numerical examples are given to demonstrate the efficiency of the proposed importance sampling scheme.

## 1. Introduction

Statistical inference for hidden Markov models has recently received some attention due to its importance in applications to speech recognition (Rabiner and Juang [24]), signal processing (Elliott *et al.* [10]), ion channel studies (Ball and Rice [3]) and molecular biology (Krogh *et al.* [16]). Good summaries on the subject are given by MacDonald and Zucchini [20], Künsch [17] and Cappé *et al.* [7]. Likelihood-based inference for hidden Markov models was first considered by Baum and Petrie [4]. Leroux [19] proved consistency of the maximum likelihood estimator (MLE) for hidden Markov chains under mild conditions. Asymptotic normality of the MLE was established by Bickel *et al.* [5].

Although asymptotic normality can be used to construct confidence regions for the parameter of interest, the lack of accuracy in the asymptotic approximation to the sampling distribution as well as the computational difficulty of the asymptotic variance–covariance matrix make it less suitable for applications. Therefore, the bootstrap method becomes a useful alternative. The application of the bootstrap method to hidden Markov models was studied by Albert [1] and Stoffer and Wall [25]. As the bootstrap estimate is obtained by Monte Carlo estimation, we need to find efficient ways to do simulation. This is particularly important for hidden Markov models, where high accuracy is often required, the estimate needs to be recomputed many times and each time a substantial amount of computation is required. For instance, when the EM (Baum–Welch) algorithm is employed to approximate the MLE, it is computed a number of times so that the error in bootstrap estimates can be assessed.

Johns [15] and Davison [8] suggested using importance sampling to construct bootstrap confidence intervals and showed that it has potential for dramatic improvement over uniform resampling. Later, Do and Hall [9] complemented it with comprehensive derivation and an empirical version. However, their method encounters difficulty in multi-parameter cases. Fuh and Hu [12] overcame the difficulty and provided an optimal tilting formula for the multi-parameter case. This helps the study of importance sampling in hidden Markov models, where the parameter space is usually multi-dimensional.

The remaining challenge is to deal with Markovian dependence. To begin, we need to determine a family of tilted distributions that contains the optimal resampling distribution, so that the optimization problem is non-trivial and solvable. Our first contribution is the construction a locally asymptotically normal (LAN) family of probability distributions around the default resampling rule. It turns out that this LAN family of distributions is closely related to the twisting formula for Markov random walks; see (A.13) in Appendix A.3. Then we minimize the asymptotic variance of the Monte Carlo estimator. Our second contribution is to provide a tilting formula for efficient importance sampling in a hidden Markov model. We also present a Poisson equation which is required to characterize the optimal tilting formula and to facilitate its implementation.

The rest of this paper is organized as follows. In Section 2 we consider a naive parametric bootstrap algorithm for hidden Markov models and importance sampling in this context. In Section 3 we propose a tilting formula for efficient importance sampling in hidden Markov models. The implementation of the formula requires a streamlined computation procedure for the variance of the associated Markov random walk. This is developed in Section 4. Numerical results are reported in Section 5. The technical details are deferred to the Appendix.

## 2. Bootstrapping hidden Markov models

### 2.1. A naive bootstrap algorithm

In this section we formulate the hidden Markov model as a Markov random walk with the underlying Markov chain as missing data. Specifically, let $\{X_t, t \geq 0\}$ be a Markov chain on a finite state space $D = \{1, 2, \ldots, d\}$, with transition probability matrix $P = [p_{ij}]_{i,j=1,\ldots,d}$, and stationary distribution $\pi = (\pi_1, \ldots, \pi_d)$. Suppose that an additive component $S_m = \sum_{t=1}^{m} Y_t$, with $Y_0 = 0$, taking values in $\mathbb{R}^\ell$, is adjoined to the chain such that $\{(X_t, S_t), t \geq 0\}$ is a Markov chain on $D \times \mathbb{R}^\ell$ and

$$P\{(X_t, S_t) \in A \times (B+y)|(X_{t-1}, S_{t-1}) = (i,y)\}$$
$$= P\{(X_t, S_t) \in A \times B|(X_{t-1}, S_{t-1}) = (i,0)\}$$
$$= P(i, A \times B) = \sum_{j \in A} \int_{y \in B} p_{ij}(\theta) f_j(y; \theta) \nu(\mathrm{d}y), \tag{2.1}$$

where $f_j(\cdot; \theta)$ is the conditional probability density function of $Y_t$ given $X_t = j$, with respect to a $\sigma$-finite measure $\nu$ on $\mathbb{R}^\ell$. Here $\theta \in \mathbb{R}^\kappa$ denotes the unknown parameter in both the transition matrix $[p_{ij}]$ and the conditional density $f_j$ of the hidden Markov model. Note that $\{X_t, t \geq 0\}$ is a Markov chain and, given $X_0, X_1, \ldots, X_m$, the random variables $Y_1, \ldots, Y_m$ are independent with density functions $f_{X_t}(\cdot; \theta)$, $t = 1, \ldots, m$.

**Definition 1.** *If there is an unobservable Markov chain $\{X_t, t \geq 0\}$ such that the process $\{(X_t, S_t), t \geq 1\}$ satisfies (2.1), then we refer to $\{S_t, t \geq 1\}$ as a hidden Markov model.*

The likelihood of a sample $\mathcal{Y} = \{y_1, \ldots, y_m\}$ from the hidden Markov model $\{S_t, t \geq 1\}$ is

$$\sum_{x_0=1}^{d} \cdots \sum_{x_m=1}^{d} \pi_{x_0} \prod_{t=1}^{m} p_{x_{t-1}, x_t} f_{x_t}(y_t; \theta), \tag{2.2}$$

where the initial distribution is the stationary distribution $\pi$. Let $\hat{\theta}$ be the MLE of $\theta$, and $V$ be an estimate of the asymptotic variance–covariance matrix of $\hat{\theta}$. We will discuss how to obtain $V$ in (3.1). Under the regularity conditions given by Bickel *et al.* ([5], pages 1617–1618), the MLE $\hat{\theta}$ is asymptotically normal. We assume these conditions hold, and henceforth refer to them together as Condition R. Let $P^{\hat{\theta}}$ be as in (2.1) such that $\theta$ equals the MLE $\hat{\theta}$ based on the observed data $\mathcal{Y}$. A bootstrap algorithm for estimating the sampling distribution of the standardized statistic $T(m) := m^{1/2} V^{-1/2}(\hat{\theta} - \theta)$ is as follows:

1. From $P^{\hat{\theta}}$, generate a Markov chain realization of $n$ steps $(x_0^*, x_1^*, \ldots, x_n^*)$.
2. For each $x_t^*$, obtain an observation $y_t^*$ by a random draw from $f_{x_t^*}(\cdot; \hat{\theta})$.
3. Compute the MLE $\hat{\theta}^*$ of the bootstrap sample $\mathcal{Y}^* = (y_1^*, \ldots, y_n^*)$ and the corresponding asymptotic variance–covariance matrix $V^*$.
4. Approximate the sampling distribution of $T(m)$ by the bootstrap distribution

$$T^*(n) = \sqrt{n}(V^*)^{-1/2}(\hat{\theta}^* - \hat{\theta}). \tag{2.3}$$

In this algorithm, we use $m$ to denote the original sample size and $n$ to denote the bootstrap sample size. We follow this notation in our discussion of efficient resampling schemes.

### 2.2. Importance sampling for bootstrap estimates

Suppose that we would like to estimate the probability of the event $\{T(m) \in A\}$ for $A \subset \mathbb{R}^\kappa$. Then the bootstrap estimate of $P\{T(m) \in A\}$ is $\hat{u} = P\{T^*(n) \in A | \mathcal{Y}\}$. Consider an importance sampling problem in hidden Markov models. Instead of resampling from $P^{\hat{\theta}}$ directly, as in the naive bootstrap algorithm described in Section 2.1, we resample from an alternative distribution $Q$. To be more precise, given $\mathcal{Y}$, let $\mathcal{Y}_1^\dagger, \ldots, \mathcal{Y}_B^\dagger$ denote independent samples drawn according to the bootstrap algorithm under the probability distribution $Q$ for the hidden Markov model $\{S_t, t \geq 1\}$. For $b = 1, \ldots, B$, write $T^b$ as the version of $T$ computed from $\mathcal{Y}_b^\dagger$. Then the importance sampling bootstrap approximation of $\hat{u}$ is

$$\hat{u}_B^\dagger = B^{-1} \sum_{b=1}^B 1_{\{T^b(n) \in A\}} \frac{\mathrm{d}P^{\hat{\theta}}}{\mathrm{d}Q}(\mathcal{Y}_b^\dagger). \tag{2.4}$$

When $Q = P^{\hat{\theta}}$, (2.4) is the approximation under the naive parametric bootstrap algorithm, the default resampling rule of this paper. To make the relationship between the default resampling rule and the importance sampling rule more transparent, we adopt the following notation. We denote the default resampling rule by replacing every occurrence of the superscript † with ∗, with the understanding that the default resampling rule is the naive parametric bootstrap algorithm. That is, $S_n^* = \sum_{t=1}^n Y_t^*$ is a hidden Markov model according to Definition 1 under the probability $P^{\hat{\theta}}$. The $P^{\hat{\theta}}$ probability is a conditional probability which depends on the sample $\mathcal{Y}$ through $\hat{\theta}$. Because we always indicate random variables from $P^{\hat{\theta}}$ with '∗', there is no danger of confusion. Henceforth, we drop the dependence on $\mathcal{Y}$ for convenience.

It is easy to see that $\hat{u}_B^\dagger$ is an unbiased estimate of $\hat{u}$. It was shown in Hall [14] that

$$\mathrm{var}(\hat{u}_B^\dagger) = B^{-1}(\hat{v} - \hat{u}^2), \qquad \text{where } \hat{v} = \mathrm{E}\left\{1_{\{T^*(n) \in A\}} \frac{\mathrm{d}P^{\hat{\theta}}}{\mathrm{d}Q}(\mathcal{Y}^*)\right\}. \tag{2.5}$$

Because $\hat{u}_B^\dagger$ is unbiased, the mean squared error of $\hat{u}_B^\dagger$ equals its variance. Note that $\hat{u}$ does not depend on $Q$. To minimize the variance (2.5) of $\hat{u}_B^\dagger$, it is sufficient to minimize $\hat{v}$ by properly choosing $Q$ from a suitable class of probability distributions.

## 3. An exponential tilting formula

### 3.1. An optimization problem in a LAN family

Under Condition R, the MLE $\hat{\theta}$ is a smooth function of the sample mean; see Ghosh ([13], Section 2.6). That is, there exists a smooth function $g$ from $\mathbb{R}^\ell \mapsto \mathbb{R}^\kappa$ such that $\hat{\theta} = g(S_m/m)$. Suppose that we would like to estimate the sampling distribution of the MLE $\hat{\theta} = (\hat{\theta}_1, \hat{\theta}_2, \ldots, \hat{\theta}_\kappa)^\mathrm{T} = (g_1(S_m/m), g_2(S_m/m), \ldots, g_\kappa(S_m/m))^\mathrm{T}$, where T denotes transpose.

Let $\Sigma$ be the $\ell \times \ell$ variance–covariance matrix of $S_m = (S_{1m}, \ldots, S_{\ell m})^{\mathrm{T}}$. Then we can estimate it by $\hat{\Sigma} = \Sigma_{\hat{\theta}}$, the variance–covariance matrix of $S_m$ under probability $P^{\hat{\theta}}$, which is assumed to be of full rank. The computation of $\hat{\Sigma}$ is discussed in Section 4. Let $\mu(\hat{\theta}) = \hat{\mu}$ be the stationary mean of $Y_t^*$ under $P^{\hat{\theta}}$. Let $J$ be the Jacobian matrix of $g$, and denote by $J_{\hat{\mu}}$ the Jacobian matrix of $g$ at $\hat{\mu}$. Let

$$V = J_{\hat{\mu}} \hat{\Sigma} J_{\hat{\mu}}^{\top} \tag{3.1}$$

be the estimated variance–covariance matrix of $\hat{\theta} = g(S_m/m)$. Note that estimating the conditional probability of the event $\{T^*(n) \in A\}$ is asymptotically equivalent to estimating

$$P\left\{(V^*)^{-1/2} J_{\hat{\mu}} \frac{S_n^* - n\hat{\mu}}{\sqrt{n}} \in A\right\}. \tag{3.2}$$

We now study the problem of how to choose $Q$ such that the variance of $\hat{u}_B^{\dagger}$ is minimized. From (2.5), this is equivalent to choosing $Q$ such that

$$\hat{v} = \min_Q \mathrm{E}\left\{1_{\{T^*(n) \in A\}} \frac{\mathrm{d}P^{\hat{\theta}}}{\mathrm{d}Q}(\mathcal{Y}^*)\right\}. \tag{3.3}$$

In order to pose (3.3) as a well-defined minimization problem, we need to determine an appropriate class of $Q$ probability distributions so that meaningful optimization can take place. It turns out that significant optimization can occur within a LAN family of probability distributions; see LeCam and Yang [18] for the definition of LAN. That is, we shall consider the family $\mathcal{C}$ of probability distributions, which are LAN at $P^{\hat{\theta}}$.

Note that in a LAN family, the magnitude of the asymptotic mean for the log-likelihood ratio equals half of its asymptotic variance. This is the key property that we need to solve the minimization problem (3.3). In Appendix A.2, we construct a LAN family and show that it is closely related to the celebrated *twisting formula* for Markov random walks, studied by Miller [21] and Ney and Nummelin [23].

When the underlying Markov chain moves from state $i$ to state $j$, we use $q_{ij}$ and $h_j(y)$ to denote respectively the transition probability and the conditional probability density of an observation $y$ under $Q$. Let $p_{ij}(\hat{\theta})$ and $f_j(y; \hat{\theta})$ be the transition probability and conditional density under $P^{\hat{\theta}}$. Note that for the rest of this section, $n \to \infty$ means the bootstrap sample size tends to infinity while the original sample size $m$ remains fixed. We now define the class $\mathcal{C}$ of probability distributions as those satisfying the following conditions:

(C1) The optimal tilting distribution is given by

$$q_{ij} h_j(y) = p_{ij}(\hat{\theta}) f_j(y; \hat{\theta}) \exp\left[-\frac{c_{ij}(y) + o(1)}{\sqrt{n}}\right], \tag{3.4}$$

(C2) Let $Z$ be a normal random variable with mean zero and variance $\sigma^2$. Then as $n \to \infty$, the log-likelihood ratio

$$L_n^* = \log\left[\frac{\mathrm{d}P^{\hat{\theta}}}{\mathrm{d}Q}(\mathcal{Y}^*)\right]$$

$$= \frac{\sum_{t=1}^n [c_{X_{t-1}^*, X_t^*}(Y_t^*) + o(1)]}{\sqrt{n}} \to Z + \frac{1}{2}\sigma^2 \quad \text{in distribution} \quad (3.5)$$

for observations $(X_t^*, Y_t^*)$, $t = 1, \ldots, n$, from $P^{\hat{\theta}}$. Moreover, $L_n^*$ and $n^{-1/2}(S_n^* - n\hat{\mu})$ are asymptotically jointly normal.

(C3) The $o(1)$ terms in (3.4) tend to 0 as $n \to \infty$ and are asymptotically negligible in determining the limiting distribution of (3.5).

For importance sampling from $Q$ in class $\mathcal{C}$, it follows from (3.3) and (3.4) that we need to minimize, over $Q$,

$$\mathrm{E}\left\{1_{\{T^*(n) \in A\}} \frac{\mathrm{d}P^{\hat{\theta}}}{\mathrm{d}Q}(\mathcal{Y}^*)\right\} = \mathrm{E}\left\{1_{\{T^*(n) \in A\}} \exp\left[\frac{\sum_{t=1}^n c_{X_{t-1}^*, X_t^*}(Y_t^*) + o(1)}{\sqrt{n}}\right]\right\}. \quad (3.6)$$

Since $T^*(n)$ is asymptotically normal and $Q \in \mathcal{C}$, it follows that as $n \to \infty$, (3.6) tends to

$$\mathrm{E}[1_{\{\mathbf{N} \in A\}} \exp(N_L)], \quad (3.7)$$

where $\mathbf{N} = (N_1, \ldots, N_\kappa)^\mathrm{T}$ and $N_L$ are jointly normal. The distribution of $\mathbf{N}$ is $\kappa$-variate normal with zero mean and identity variance–covariance matrix, while the distribution of $N_L$ is normal with mean $\mu_L$ and variance $\sigma_L^2$. By (3.5), we have $\mu_L = \sigma_L^2/2$.

Let $\rho_k$ be the asymptotic correlation between the $k$th component, $\sqrt{n}(\hat{\theta}_k^* - \hat{\theta}_k)$, of $T^*(n)$ and the log-likelihood ratio $L_n^*$ for $k = 1, \ldots, \kappa$. Let $\Sigma_L = (\sigma_L \rho_1, \sigma_L \rho_2, \ldots, \sigma_L \rho_\kappa)^\mathrm{T}$ denote the covariance between $\mathbf{N}$ and $L_n^*$. Then we can write the joint variance–covariance matrix of $(\mathbf{N}, N_L)^\mathrm{T}$ as

$$\begin{pmatrix} 1 & 0 & \cdots & 0 & \rho_1 \sigma_L \\ 0 & 1 & \ddots & \vdots & \rho_2 \sigma_L \\ \vdots & \ddots & \ddots & 0 & \vdots \\ 0 & \cdots & 0 & 1 & \rho_\kappa \sigma_L \\ \rho_1 \sigma_L & \cdots & \cdots & \rho_\kappa \sigma_L & \sigma_L^2 \end{pmatrix} = \begin{pmatrix} \mathbf{I}_\kappa & \Sigma_L \\ \Sigma_L^\mathrm{T} & \sigma_L^2 \end{pmatrix},$$

where $\mathbf{I}_\kappa$ is the $\kappa \times \kappa$ identity matrix. Thus the optimization problem (3.3) is reduced to that of finding $\sigma_L$ and $\Sigma_L$ so that (3.7) is minimized.

### 3.2. The derivation of the optimal tilting formula

The following lemma determines the minimum of (3.7). The proof of Lemma 1 can be found in Fuh and Hu [12].

**Lemma 1.** *The following choice of $\Sigma_L$ and $\sigma_L$ minimizes* (3.7):

$$\Sigma_L = -\tfrac{1}{2}\mathrm{E}(\mathbf{N}|\mathbf{N} \in A - \Sigma_L), \qquad \sigma_L = \sqrt{\Sigma_L^{\mathrm{T}}\Sigma_L}. \tag{3.8}$$

We now proceed to identify $c_{ij}(y)$ in (3.4) for the optimal $Q$ such that (3.8) is satisfied. Observe that $\lim_{n\to\infty} \mathrm{cov}(\Sigma_L^{\mathrm{T}} T^*(n), L_n^*) = \lim_{n\to\infty} \Sigma_L^{\mathrm{T}} \mathrm{cov}(T^*(n), L_n^*) = \Sigma_L^{\mathrm{T}} \Sigma_L = \sigma_L^2$. On the other hand, from the Cauchy–Schwarz inequality it follows that

$$\lim_{n\to\infty} \mathrm{cov}(\Sigma_L^{\mathrm{T}} T^*(n), L_n^*) \le \lim_{n\to\infty} \sqrt{\mathrm{var}(\Sigma_L^{\mathrm{T}} T^*(n))\mathrm{var}(L_n^*)} = \sigma_L^2.$$

Because the equality is attained only when $L_n^*$ is asymptotically equivalent to a linear function of $\Sigma_L^{\mathrm{T}} T^*(n)$ and thus asymptotically equivalent to a linear function of $S_n$, we have

$$n^{-1/2} \sum_{t=1}^n [c_{X_{t-1}^*, X_t^*}(Y_t^*) + o(1)] \approx c n^{-1/2}(S_n^* - n\hat\mu) \tag{3.9}$$

for some constant $c \in \mathbb{R}^\ell$.

Let $n_{ij}$ be the number of $i$-to-$j$ transitions and $n_i$ be the number of visits to state $i$ by $X_1^*, \ldots, X_n^*$. We first represent $n_i/\sqrt{n}$ in terms of $n_{ij}/\sqrt{n}$. Let $\gamma_i$ be any constant independent of $n$; then $n_i n^{-1/2} \approx n^{-1/2}[\gamma_i \sum_{j=1}^d n_{ij} + (1-\gamma_i)\sum_{j=1}^d n_{ji}]$, as $n \to \infty$. This is possible because $\sum_{j=1}^d n_{ij} - \sum_{j=1}^d n_{ji} = 1_i(X_1^*) - 1_i(X_n^*)$, where $1_i(\cdot)$ denotes the indicator function of state $i$. Later, we will specify the value of $\gamma_i$ so that other conditions are satisfied. Let us first assume that $\ell = 1$ – that is, $Y_t^*, t = 1, 2, \ldots, n$, are one-dimensional – and then show that the generalization to the multi-dimensional case is straightforward. Let $w_i = \sum_{t \in D_i} Y_t^*/n_i$, $i = 1, \ldots, d$, where $D_i = \{t|X_t^* = i,\ 1 \le t \le n\}$.

For $i = 1, \ldots, d$, summing $w_i - \hat\mu$ with respect to $i$ from 1 to $d$, we obtain

$$c\frac{S_n^* - n\hat\mu}{\sqrt{n}} = c\sum_{i=1}^d (w_i - \hat\mu)\frac{n_i}{\sqrt{n}} \approx c\sum_{i=1}^d \left[(w_i - \hat\mu)\gamma_i \sum_{j=1}^d \frac{n_{ij}}{\sqrt{n}} + (w_i - \hat\mu)(1-\gamma_i)\sum_{j=1}^d \frac{n_{ji}}{\sqrt{n}}\right]$$

$$= c\sum_{i=1}^d (w_i - \hat\mu)\frac{n_{ii}}{\sqrt{n}} + c\sum_{i,j=1, i\ne j}^d (w_j - \hat\mu - \delta_i + \delta_j)\frac{n_{ij}}{\sqrt{n}}, \tag{3.10}$$

where $\delta_i = -(w_i - \hat\mu)\gamma_i$. Let $D_{ij} = \{t|X_{t-1}^* = i, X_t^* = j, 2 \le t \le n\}$. By (3.9), match the coefficient of (3.10) with $\sum_{i,j=1}^d \bar c_{ij} n_{ij}/\sqrt{n}$, where $\bar c_{ij} = \sum_{t \in D_{ij}} c_{ij}(Y_t^*)/n_{ij}$, to obtain

$$\lim_{n\to\infty} \bar c_{ij} = \lim_{n\to\infty} c(w_j - \hat\mu - \delta_i + \delta_j). \tag{3.11}$$

Note that $\lim_{n\to\infty} \bar c_{ij} = \int c_{ij}(y) f_j(y, \hat\theta)\nu(\mathrm{d}y)$. In view of

$$1 = \sum_{j=1}^d \int \exp\left[-\frac{c_{ij}(y) + o(1)}{n^{1/2}}\right] f_j(y, \hat\theta) p_{ij}(\hat\theta)\nu(\mathrm{d}y)$$

$$= \sum_{j=1}^{d} \int [1 - c_{ij}(y)n^{-1/2} + o(n^{-1/2})] f_j(y,\hat{\theta}) p_{ij}(\hat{\theta}) \nu(dy),$$

we have

$$\lim_{n\to\infty} \sum_{j=1}^{d} \bar{c}_{ij} p_{ij}(\hat{\theta}) = 0. \tag{3.12}$$

From (3.11) and (3.12), we conclude that $\delta_i$, $i = 1, \ldots, d$, satisfy

$$\lim_{n\to\infty} \sum_{j=1}^{d}(w_j - \hat{\mu} - \delta_i + \delta_j)p_{ij}(\hat{\theta}) = 0 \quad \Rightarrow \quad \lim_{n\to\infty} \sum_{j=1}^{d}(w_j - \hat{\mu})p_{ij}(\hat{\theta}) - \delta_i + \sum_{j=1}^{d} \delta_j p_{ij}(\hat{\theta}) = 0.$$

In matrix form, this becomes

$$\lim_{n\to\infty} \begin{bmatrix} p_{11} & p_{12} & \cdots & p_{1d} \\ p_{21} & p_{22} & \cdots & p_{2d} \\ \vdots & \vdots & \ddots & \vdots \\ p_{d1} & p_{d2} & \cdots & p_{dd} \end{bmatrix} \begin{bmatrix} (w_1 - \hat{\mu}) \\ (w_2 - \hat{\mu}) \\ \vdots \\ (w_d - \hat{\mu}) \end{bmatrix}$$
$$- \begin{bmatrix} 1 - p_{11} & -p_{12} & \cdots & -p_{1d} \\ -p_{21} & 1 - p_{22} & \cdots & -p_{2d} \\ \vdots & \vdots & \ddots & \vdots \\ -p_{d1} & -p_{d2} & \cdots & 1 - p_{dd} \end{bmatrix} \begin{bmatrix} \delta_1 \\ \delta_2 \\ \vdots \\ \delta_d \end{bmatrix} = 0,$$

where we have dropped $\hat{\theta}$ from $p_{ij}(\hat{\theta})$ for simplicity. Clearly, $\lim_{n\to\infty} w_j = E(Y_t^* | X_t^* = j) = \mu_j$. Thus we can replace $w_j$ by $\mu_j$ in the preceding matrix equation.

It is easy to see that if $Y_t^*$, $t = 1, \ldots, n$, are multi-dimensional, the preceding matrix equality holds for each component of the random vectors $Y_t^*$. Denote by $I$ as the identity matrix, and let $\Gamma_i = E(Y_t^* - \hat{\mu} | X_t^* = i)$ be the adjusted conditional mean given $X_t^* = i$, for $i = 1, \ldots, d$. Write $\Gamma = (\Gamma_i)$ and $\Delta = (\delta_i) = (\delta_{il})$, a $d \times \ell$ matrix. Then the preceding matrix equation implies that $\Delta$ is a solution of the Poisson equation

$$(I - P)\Delta = P\Gamma. \tag{3.13}$$

Let $\delta_i$ be the solution of (3.13). Consider choosing

$$c_{ij}(y) = \Sigma_L^T V^{-1/2} J_{\hat{\mu}}(y - \hat{\mu} + \delta_i - \delta_j), \tag{3.14}$$

where $\Sigma_L$ is defined in (3.8). It can be shown that if we choose $c_{ij}(y)$ according to (3.14), then (3.8) is satisfied.

The optimal tilting distribution is given by

$$q_{ij} h_j(y) = \frac{p_{ij}(\hat{\theta}) f_j(y, \hat{\theta}) \exp[-c_{ij}(y)/\sqrt{n}]}{\sum_{j=1}^{d} p_{ij}(\hat{\theta}) \int \exp(-c_{ij}(y)/\sqrt{n}) f_j(y, \hat{\theta}) \nu(dy)}, \tag{3.15}$$

where $c_{ij}(y)$ given by (3.14). Furthermore, let

$$C_{ij} = \int \exp[-c_{ij}(y)/\sqrt{n}] f_j(y,\hat{\theta}) \nu(\mathrm{d}y).$$

Then, in (3.15), we have

$$q_{ij} = \frac{p_{ij}(\hat{\theta}) C_{ij}}{\sum_{j=1}^{d} p_{ij}(\hat{\theta}) C_{ij}} \tag{3.16}$$

and

$$h_j(y) = C_{ij}^{-1} \exp(-c_{ij}(y)/\sqrt{n}) f_j(y,\hat{\theta}). \tag{3.17}$$

Note that due to the cancelation of $\delta_i$ and $\delta_j$ in $c_{ij}(y)$ and $C_{ij}^{-1}$, $h_j(y)$ defined in (3.17) depends on the current state $j$ only. We summarize our findings in the following theorem.

**Theorem 1.** *Let $\hat{\theta}$ be the MLE of the sample $\mathcal{Y} = \{y_1, y_2, \ldots, y_m\}$ from a hidden Markov model* (2.1) *satisfying Condition R. To estimate the sampling distribution of $\hat{\theta}$, we do importance sampling according to the following procedure.*

(i) *Sample from a Markov chain with transition matrix* (3.16) *to obtain $\{x_0^\dagger, x_1^\dagger, \ldots, x_n^\dagger\}$.*
(ii) *For each $x_i^\dagger$, $i = 1, \ldots, n$, sample from $h_{x_i^\dagger}(\cdot)$ of* (3.17) *to obtain $y_i^\dagger$.*
(iii) *Calculate the MLE $\hat{\theta}^\dagger$ of the sample $\{y_1^\dagger, \ldots, y_n^\dagger\}$.*
(iv) *Repeat the preceding steps B times to obtain an approximation of* (3.2) *via* (2.4).

*The preceding importance sampling scheme minimizes the asymptotic variance of* (2.4) *among all distributions within the class $\mathcal{C}$ defined in* (3.4).

**Proof.** We have shown that in order for the importance sampling estimator (2.4) to have minimum asymptotic variance in class $\mathcal{C}$, it is necessary for the importance sampling distribution to satisfy (3.14)–(3.17). It remains to show that the resampling distribution given by (3.14)–(3.17) actually belongs to $\mathcal{C}$. The details are given in Appendix A.1. □

## 4. Implementation of the tilting formula

To implement the tilting formula (3.14)–(3.17), we need to know how to compute $\Sigma_L$ and $V$. Let us first consider the computation of $\Sigma_L$. Since $\Sigma_L$ is only implicitly defined by (3.8), it cannot be evaluated directly. However, it can be employed to construct a recursive approximation algorithm. In this regard, it is easier to approximate $-\Sigma_L$. Changing $\Sigma_L$ to $\bar{\Sigma}_L = -\Sigma_L$ in (3.8), we obtain

$$\bar{\Sigma}_L = \tfrac{1}{2} \mathrm{E}(\mathbf{N} | \mathbf{N} \in A + \bar{\Sigma}_L). \tag{4.1}$$

From (4.1), we can compute $\bar{\Sigma}_L$ via a recursive algorithm as follows:

(i) Initialize $\bar{\Sigma}_L = \bar{\Sigma}_L^{(0)}$.
(ii) Iterate $\bar{\Sigma}_L^{(i+1)} = \frac{1}{2}E[\mathbf{N}|\mathbf{N} \in A + \bar{\Sigma}_L^{(i)}]$.

The convergence proof and some useful results on the implementation of the recursive algorithm can be found in Fuh and Hu ([12], Section 2).

The computation of $V$, or, for that matter, the computation of $\Sigma$, the asymptotical variance for the Markov random walk $S_n$, is much more complicated than that for the independent and identically distributed case where the sample covariance matrix does the job. Here we develop a representation which allows straightforward calculation of $\Sigma$ via the solution of a Poisson equation.

Let $\{X_n, n \geq 0\}$ be a finite ergodic (positive recurrent, aperiodic and irreducible) Markov chain on state space $D = \{1, \ldots, d\}$ with stationary distribution $\pi$. Let $\{(X_n, S_n), n \geq 0\}$ be the Markov random walk defined in (2.1). Then

$$\Sigma = \mathrm{E}_\pi[(Y_1 - \mu)(Y_1 - \mu)^\mathrm{T})] + 2\sum_{k=1}^{\infty} \mathrm{E}_\pi[(Y_1 - \mu)(Y_{k+1} - \mu)^\mathrm{T}] \qquad (4.2)$$

is well defined if $\mathrm{E}_\pi(\|Y_1\|^2) < \infty$. Furthermore, if $\Sigma$ is positive definite, Theorem 17.0.1 of Meyn and Tweedie [22] shows that

$$\frac{1}{\sqrt{n}}(S_n - n\mu) \to N(0, \Sigma) \qquad \text{in distribution.} \qquad (4.3)$$

Note that (4.2) is inconvenient to compute. We provide another representation of the asymptotic variance, $\Sigma = [\sigma_{ll'}^2]_{l,l'=1,\ldots,\ell}$, which facilitates the computation of it,

$$\sigma_{ll'}^2 = \sum_{i=1}^{d}[G_{ll'}(i) - \Gamma_l(i)\Gamma_{l'}(i)]\pi_i + \sum_{i,j=1}^{d}[\Gamma_l(j) - \delta_{il} + \delta_{jl}]^2 p_{ij}\pi_i, \qquad (4.4)$$

where $\Gamma_l(i) = \mathrm{E}(Y_{1l} - \mu_l|X_0 = i), G_{ll'}(i) = \mathrm{E}[(Y_{1l} - \mu_l)(Y_{1l'} - \mu_{l'})^\mathrm{T}|X_0 = i]$, $\mu_l$ is the $l$th component of the stationary mean, and $\delta_{il}, i = 1, \ldots, d, l = 1, \ldots, \ell$, are the elements of the $d \times \ell$ matrix $\Delta$ which is the solution of the Poisson equation

$$(I - P)\Delta = P\Gamma_l(i), \qquad (4.5)$$

in which $I$ denotes the identity operator. The asymptotic variance formulae (4.4) and (4.5) for Markov random walks in general state space and their proofs are given in Appendix A.2.

## 5. Simulation study

To demonstrate the effectiveness of our method, we study two examples. We measure the effectiveness by relative efficiency, which is defined to be the ratio of the variance under the default resampling distribution $P$ to that under the tilted probability distribution $Q$ given by (3.14)–(3.17). We refer the reader to Fuh and Hu ([12], Section 5.1) for results on relative efficiency.

To construct confidence regions through importance sampling, it is usually necessary to combine stratified sampling with importance sampling. That is, we need to partition the region into several parts and apply a different tilting formula to each part. These results are given in Fuh and Hu ([12], Section 4) and are used in Sections 5.1 and 5.2.

The first example concerns a three-state hidden Markov chain $\{X_t, t \geq 0\}$ with transition matrix $P$, and the $\{Y_t, t \geq 0\}$ observations follow a bivariate normal distribution $N((\mu_{j1}, \mu_{j2})^{\mathrm{T}}, \Sigma)$ given the states $j = 1, 2, 3$ of the Markov chain. Specifically, let

$$P = \begin{bmatrix} 0.2 & 0.3 & 0.5 \\ 0.3 & 0.4 & 0.3 \\ 0.5 & 0.3 & 0.2 \end{bmatrix}, \qquad \Sigma = \begin{bmatrix} 1.0 & 0.3 \\ 0.3 & 1.0 \end{bmatrix}, \tag{5.1}$$

$\mu_{11} = \mu_{12} = 0$, $\mu_{21} = \mu_{22} = 5$ and $\mu_{31} = \mu_{32} = 10$. The second example concerns the time series of daily counts of epileptic seizures is given in Section 5.2.

## 5.1. A bivariate normal example

We first generate $m = 100$ observations $Y_1 = (Y_{11}, Y_{12})^{\mathrm{T}}, \ldots, Y_m = (Y_{m1}, Y_{m2})^{\mathrm{T}}$ from the hidden Markov model defined in (5.1). The parameter of interest is the stationary mean $(\sum_{i=1}^{3} \pi_i \mu_{i1}, \sum_{i=1}^{3} \pi_i \mu_{i2})^{\mathrm{T}}$. The estimator is the sample mean $(\hat{\mu}_1, \hat{\mu}_2)^{\mathrm{T}}$. Two different types of confidence regions, square and circular, are considered in this simulation study. The bootstrap sample size is $n = 100$, and the number of bootstrap replications is $B = 1000$ for uniform resampling and $B = 200$, 100, 50 for importance sampling. The whole experiment was repeated 180 000 times to estimate the coverage probability and the mean and the standard deviation of confidence region areas. The nominal coverage probability is 0.95 in all cases. The results are summarized in Tables 1 and 2. The tilting points and the confidence regions are shown in Figures 1 and 2 for various nominal levels.

In Table 1, we divide the complement of a square region into four subregions. The four optimal tilting points chosen according to (3.14) and (3.15) are $(0, r)$, $(r, 0)$, $(0, -r)$, $(-r, 0)$, as shown in Figure 1(a), where $r = 2.4613$ by inverting the normal approximation. Note that the large-deviations tilting would use $r = 2.236$; see Fuh and Hu [12]. The four transition matrices and the conditional densities $h_j^k(\cdot) \sim N((\mu_{j1}^k, \mu_{j2}^k)^{\mathrm{T}}, \Sigma)$ of $Y$ given the hidden states $j = 1, 2, 3$ and optimal tilting points $k = 1, \ldots, 4$ can be calculated from (3.14)–(3.17).

In Table 2, we divide the complement of a circular region into four subregions. The four optimal tilting points chosen according to (3.8) are $(0, r)$, $(r, 0)$, $(0, -r)$, $(-r, 0)$, as shown in Figure 2(a), where $r = 2.655$, whereas the large-deviations tilting would use $r = 2.447$. Similar to the square confidence region, the transition matrices and conditional densities can be calculated from (3.14)–(3.17).

Tables 1 and 2 reveal that the importance sampling method permits a reduction of replication sizes from 5 to 1. The performance is still reasonable for a reduction from 10 to 1. The only penalty seems to be a slight increase in the variability of the confidence region area.

## 5.2. A Poisson example

Albert [1] described the fitting of a two-state Poisson hidden Markov model to the sequence of daily seizure counts recorded during follow-up for each of 13 outpatients with intractable epilepsy maintained on steady anticonvulsant drugs. Specifically, let $\mathbf{X} = (X_0, X_1, \ldots, X_m)$ be generated from a two-state (0 and 1) Markov chain with unknown transition probabilities $p_{01}$ and $p_{10}$. Write $p_{11} = 1 - p_{10}$ and $p_{00} = 1 - p_{01}$. Given

**Table 1.** Square confidence region

| Bootstrap method | Bootstrap replication size | Non-coverage probability | Region area | |
|---|---|---|---|---|
| | | | Average | Standard deviation |
| Non-studentized statistic | | | | |
| Ordinary | 1000 | 0.0493 | 2.396 | 0.644 |
| Tilted | 200 | 0.0504 | 2.394 | 0.647 |
| Tilted | 100 | 0.0484 | 2.407 | 0.651 |
| Tilted | 52 | 0.0475 | 2.421 | 0.657 |
| Studentized statistic | | | | |
| Ordinary | 1000 | 0.0495 | 2.394 | 0.641 |
| Tilted | 200 | 0.0502 | 2.393 | 0.640 |
| Tilted | 100 | 0.0490 | 2.403 | 0.642 |
| Tilted | 52 | 0.0485 | 2.409 | 0.645 |

**Table 2.** Circular confidence region

| Bootstrap method | Bootstrap replication size | Non-coverage probability | Region area | |
|---|---|---|---|---|
| | | | Average | Standard deviation |
| Nonstudentized statistic | | | | |
| Ordinary | 1000 | 0.0515 | 2.383 | 0.644 |
| Tilted | 200 | 0.0491 | 2.393 | 0.645 |
| Tilted | 100 | 0.0485 | 2.399 | 0.653 |
| Tilted | 52 | 0.0467 | 2.414 | 0.659 |
| Studentized statistic | | | | |
| Ordinary | 1000 | 0.0510 | 2.391 | 0.641 |
| Tilted | 200 | 0.0492 | 2.392 | 0.641 |
| Tilted | 100 | 0.0489 | 2.394 | 0.643 |
| Tilted | 52 | 0.0472 | 2.409 | 0.645 |

**X**, let $Y_1, \ldots, Y_m$ be the observed counts from the Poisson distributions

$$P(Y_k = y_k | X_k = i) = \frac{e^{-\lambda_i} \lambda_i^{y_k}}{y_k!}, \qquad i = 0, 1,$$

where $\lambda_0$ and $\lambda_1$ are the mean numbers of counts in states 0 and 1, respectively. Let $\theta = (p_{01}, p_{10}, \lambda_0, \lambda_1)$ be the parameter of interest. Balish *et al.* [2] demonstrated using quasi-likelihood regression models that all but one patient had seizure counts fitted inadequately by a Poisson distribution. As reported in Albert [1], the two-state hidden Markov model provides a better fit and described the apparent clustering of seizures better than a Poisson regression model with autoregressive terms.

To illustrate the efficiency of the proposed method in Theorem 1, we adopted estimates in Albert [1] of transition probabilities $\hat{p}_{01} = 0.197$, $\hat{p}_{10} = 0.61$, and the Poisson means $\hat{\lambda}_0 = 0.251, \hat{\lambda}_1 = 2.0$ for a particular patient. Bootstraps are done by generating a random sample of size 100 with the aforementioned parameter values for the patient concerned. We then compute the MLE using the EM algorithm and generate bootstrap samples via the naive bootstrap algorithm in Section 2.1 and via importance sampling according to Theorem 1. The number of bootstrap replications is $B = 1000$ for uniform resampling, and $B = 200$, 100, 52 are used for importance sampling. As in Section 5.1, we obtain four tilting points and $r = 2.4613$. Table 3 shows that importance sampling permits a reduction of bootstrap replication sizes from 5 to 1. Figure 3 exhibits three confidence regions for $(\lambda_0, \lambda_1)$ from importance sampling with $n = 100$ and $B = 200$.

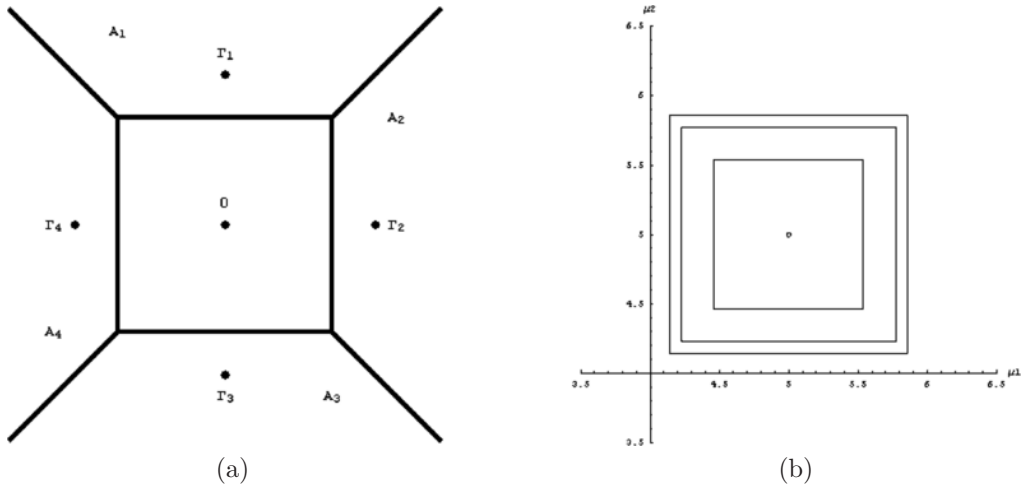

**Figure 1.** (a) The four tilting points and (b) 0.5, 0.95 and 0.99 bootstrap confidence regions for parameter estimates of a three-state model using importance sampling with $n = 100$ and replication size $B = 200$.

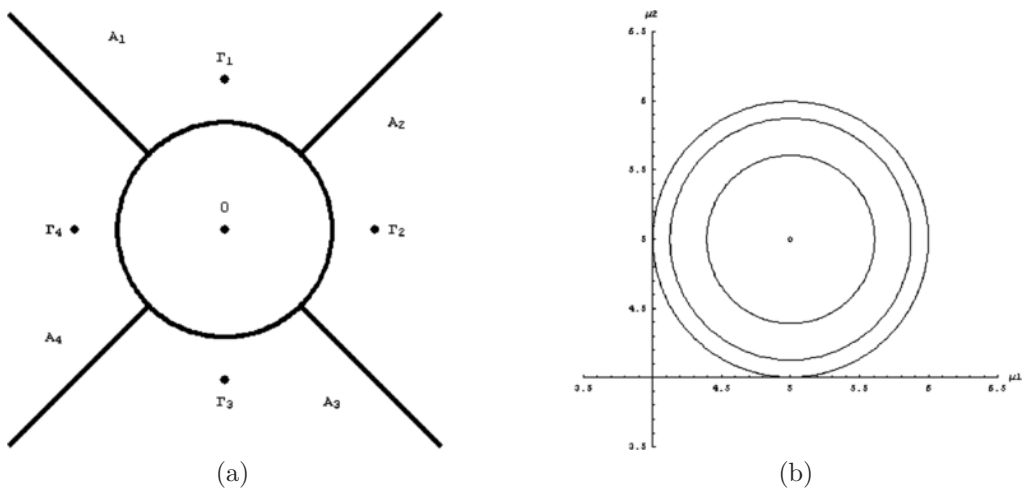

**Figure 2.** (a) The four tilting points and (b) the 0.5, 0.95 and 0.99 bootstrap circular confidence regions for parameter estimates of a three-state hidden Markov model using importance sampling with $n = 100$ and replication size $B = 200$.

**Table 3.** Square confidence regions for the two-state model with parameters $\lambda_0 = 0.251$, $\lambda_1 = 2.0$, $p_{01} = 0.197$, $p_{10} = 0.61$ and 10 000 Monte Carlo repetitions

| Bootstrap method | Replication size | Region area Average | S.D. |
|---|---|---|---|
| Ordinary | 1000 | 5.761 | 1.674 |
| Tilted | 200 | 5.789 | 1.656 |
| Tilted | 100 | 6.069 | 1.702 |
| Tilted | 52 | 6.446 | 1.761 |

## Appendix

### A.1. Proof of Theorem 1

We drop '*' from $y_t^*$ and $x_t^*$ for simplicity. It is understood that in the following proof $y_t$ and $x_t$ are generated according to $P^{\hat{\theta}}$, that is, the default sampling rule. From (3.15), the log-likelihood ratio is given by

$$\log \prod_{t=1}^{n} \frac{p_{x_{t-1},x_t} f_{x_t}(y_t, \hat{\theta})}{q_{x_{t-1},x_t} h_{x_t}(y_t)}$$

$$= \log \prod_{t=1}^{n} \sum_{j=1}^{d} p_{x_{t-1},j} \int \exp\left[-\frac{c_{x_{t-1},j}(y_t)}{n^{1/2}}\right] f_j(y_t,\hat{\theta})\nu(\mathrm{d}y) \exp\left[\frac{c_{x_{t-1},x_t}(y_t)}{n^{1/2}}\right]$$

$$= \sum_{t=1}^{n} \left\{ \log\left[\sum_{j=1}^{d} p_{x_{t-1},j} \int \exp\left(-\frac{c_{x_{t-1},j}(y_t)}{n^{1/2}}\right) f_j(y_t,\hat{\theta})\nu(\mathrm{d}y)\right] + \frac{c_{x_{t-1},x_t}(y_t)}{n^{1/2}} \right\}. \quad (A.1)$$

Let $E_j$ denote the conditional expectation given $x_t = j$. The integral in (A.1) with a two-term Taylor expansion of the exponential term equals

$$1 - \frac{E_j[c_{x_{t-1},j}(y_t)]}{n^{1/2}} + \frac{E_j[c^2_{x_{t-1},j}(y_t)]}{2n} + o(n^{-1}).$$

By (3.12), multiplying the preceding expression by $p_{x_{t-1},j}$, summing over $j$, and taking logarithms yield

$$\log\left[1 + \sum_{j=1}^{d} \frac{E_j[c^2_{x_{t-1},j}(y_t)]p_{x_{t-1},j}}{2n} + o(n^{-1})\right] \approx \sum_{j=1}^{d} \frac{E_j[c^2_{x_{t-1},j}(y_t)]p_{x_{t-1},j}}{2n}.$$

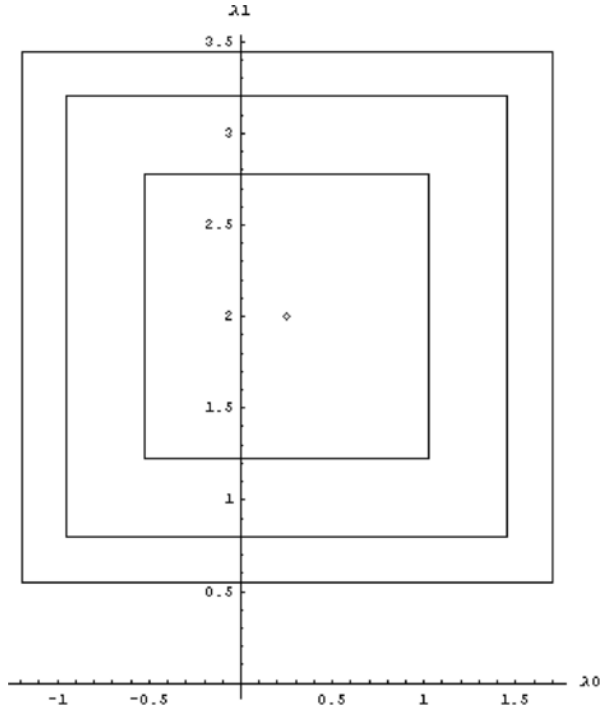

**Figure 3.** The 0.5, 0.95 and 0.99 confidence regions with parameters $\lambda_0 = 0.251$, $\lambda_1 = 2.0$, $p_{01} = 0.197$, $p_{10} = 0.61$ using importance sampling with $n = 100$ and replication $B = 200$.

In view of (3.14), summing the preceding expression over $t$, we find that the log term in (A.1) is approximated by

$$\sum_{t=1}^{n} \Sigma_L^T \Sigma_{\hat{\theta}}^{-1/2} J_{\hat{\mu}}^T \sum_{j=1}^{d} \left[ \frac{E_j(y_t - \hat{\mu} - \delta_{x_{t-1}} + \delta_j)(y_t - \hat{\mu} - \delta_{x_{t-1}} + \delta_j)^T p_{x_{t-1},j}}{2n} \right] J_{\hat{\mu}} \Sigma_{\hat{\theta}}^{-1/2} \Sigma_L.$$

When $n$ is large, the proportion of time that the chain $\{x_t\}$ spends in state $i$ is approximately $\pi_i$, the stationary probability of state $i$. Using this fact in the preceding expression, we see that it is approximated by

$$\tfrac{1}{2}\Sigma_L^T \Sigma_{\hat{\theta}}^{-1/2} J_{\hat{\mu}}^T \sum_{i=1}^{d}\sum_{j=1}^{d} [E_j(y_t - \hat{\mu} - \delta_i + \delta_j)(y_t - \hat{\mu} - \delta_i + \delta_j)^T p_{ij}\pi_i] J_{\hat{\mu}} \Sigma_{\hat{\theta}}^{-1/2} \Sigma_L.$$

The summation over $i,j$ of the terms in the square bracket equals

$$\sum_{j=1}^{d} E_j(y_t - \mu_j)(y_t - \mu_j)^T \pi_j + \sum_{i=1}^{d}\sum_{j=1}^{d} [E_j(y_t - \hat{\mu} - \delta_i + \delta_j)][E_j(y_t - \hat{\mu} - \delta_i + \delta_j)]^T p_{ij}\pi_i.$$

The first term of the preceding expression can be rewritten as

$$\sum_{j=1}^{d} E_j[(y_t - \hat{\mu})(y_t - \hat{\mu})^T - (\mu_j - \hat{\mu})(\mu_j - \hat{\mu})^T]\pi_j,$$

and adding the second term shows that the sum is identical to the variance given by (4.4). By (3.1), we conclude that the log term in (A.1) equals $\tfrac{1}{2}\Sigma_L^T \Sigma_L = \tfrac{1}{2}\sigma_L^2$ asymptotically.

Consider the last term in (A.1). By (3.14), we have

$$\sum_{t=1}^{n} \frac{c_{x_{t-1},x_t}(y_t)}{n^{1/2}} = \sum_{t=1}^{n} \Sigma_L^T \Sigma_{\hat{\theta}}^{-1/2} J_{\hat{\mu}}^T \frac{y_t - \hat{\mu} - \delta_{x_{t-1}} + \delta_{x_t}}{n^{1/2}} = \Sigma_L^T \Sigma_{\hat{\theta}}^{-1/2} J_{\hat{\mu}}^T \frac{s_n - n\hat{\mu} - \delta_{x_1} + \delta_{x_n}}{n^{1/2}}.$$

Note that the last two terms in the numerator are negligible after dividing by $n^{1/2}$. By (4.3), the preceding expression converges, in distribution, to a normal random variable with mean zero and variance $\sigma_L^2$. This shows that (3.5) is satisfied, which completes the proof.

## A.2. Asymptotic variance of Markov random walks

Let $\{X_n, n \geq 0\}$ be an aperiodic and irreducible Markov chain on a general state space $D$ with $\sigma$-algebra $\mathcal{D}$. The irreducibility is with respect to a maximal irreducibility measure $\varphi$ on $\mathcal{D}$; see Meyn and Tweedie ([22], page 89) for definition. Suppose that an additive component $S_n = \sum_{k=0}^{n} Y_k$ with $S_0 = Y_0 = 0$, taking values in $\mathbb{R}^\ell$, is adjoined to the chain such that $\{(X_n, S_n),\ n \geq 0\}$ is a Markov chain on $D \times \mathbb{R}^\ell$ and

$$P\{(X_{n+1}, S_{n+1}) \in A \times (B+s) | (X_n, S_n) = (x, s)\}$$
$$= P\{(X_1, S_1) \in A \times B | (X_0, S_0) = (x, 0)\} = P(x, A \times B)$$

for all $x \in D$, $s \in \mathbb{R}^\ell$, $A \in \mathcal{D}$ and $B \in \mathcal{B}(\mathbb{R}^\ell)$, the Borel $\sigma$-algebra of $\mathbb{R}^\ell$. The chain $\{(X_n, S_n), n \geq 0\}$ is referred to as a *Markov additive process*, and its additive component $S_n$ as a *Markov random walk*.

Let $\nu$ be an initial distribution on $X_0$ and let $\mathrm{E}_\nu$ and $\mathrm{var}_\nu$ denote expectation and variance under $\nu$, respectively. If $\nu$ is degenerate at $x$, we simply write $\mathrm{E}_x(\mathrm{var}_x)$. If $\{X_n, n \geq 0\}$ has a unique stationary measure $\pi$, let $\mu := \int_D \mathrm{E}_x(\xi_1) \pi(\mathrm{d}x)$ denote the stationary mean.

For any real-valued non-negative kernel $\{K(x, A); x \in D, A \in \mathcal{D}\}$, function $h : D \to \mathbb{R}$, and measure $\Psi$ on $(D, \mathcal{D})$, write

$$Kh(x) = \int_D K(x, \mathrm{d}y) h(y), \qquad \Psi K(A) = \int_D \Psi(\mathrm{d}x) K(x, A),$$
$$\Psi h(A) = \int_A \Psi(\mathrm{d}x) h(x) \quad \text{(a sign measure)}, \tag{A.2}$$
$$\Psi h = \Psi h(D) \qquad \text{(a real number)}.$$

***Condition K1*** *Minorization. There exist a probability measure $\Psi$ on $D \times \mathbb{R}^\ell$ and a measurable function $h$ on $D$ such that $\int h(x) \pi(\mathrm{d}x) > 0$, $\int \Psi(\mathrm{d}x \times \mathbb{R}^\ell) h(x) > 0$, and*

$$P(x, A \times B) \geq h(x) \Psi(A \times B), \tag{A.3}$$

*for all $x \in D$, $A \in \mathcal{D}$, $B \in \mathcal{B}(\mathbb{R}^\ell)$.*

For an aperiodic and irreducible Markov random walk satisfying Condition K1, by making use of a splitting chain argument, there exists an equivalent Markov chain with a recurrent state; see, for example, Meyn and Tweedie ([22], Chapter 5). Thus, without loss of generality, we assume that there exists a recurrent state $\Delta$ in $D$ such that the Markov chain $X_n$ visits the state $\Delta$ infinitely often. Let $T_\Delta = \inf\{n \geq 1 : X_n = \Delta\}$ be the first recurrent time.

**Theorem 2.** *Let $\{(X_n, S_n), n \geq 0\}$ be a Markov random walk on state space $D$, satisfying Condition K1. Assume*

$$\mathrm{E}_\Delta \left( \sum_{k=1}^{T_\Delta - 1} \|Y_k\|^2 \right) < \infty \quad \text{and} \quad \mathrm{E}_\Delta T_\Delta^2 < \infty. \tag{A.4}$$

*Then*

$$\Sigma := \mathrm{E}_\pi[(Y_1 - \mu)(Y_1 - \mu)^\mathrm{T}] + 2\sum_{k=1}^\infty \mathrm{E}_\pi[(Y_1 - \mu)(Y_{k+1} - \mu)^\mathrm{T}] \quad \text{(A.5)}$$

*is well defined. Furthermore, if $\Sigma$ is positive definite, then*

$$\frac{1}{\sqrt{n}}(S_n - n\mu) \longrightarrow N(0, \Sigma) \qquad \text{in distribution.} \quad \text{(A.6)}$$

*The asymptotic variance, $\Sigma = [\sigma_{ll'}^2]_{l,l'=1,\ldots,\ell}$, can be calculated via*

$$\int_D [G_{ll'} - \Gamma_l \Gamma_{l'}]\pi(\mathrm{d}x) + \int_D [\Gamma_l(x') - \delta_{xl} + \delta_{x'l}][\Gamma_{l'}(x') - \delta_{xl'} + \delta_{x'l'}]P(x, \mathrm{d}x')\pi(\mathrm{d}x), \quad \text{(A.7)}$$

*where $\Gamma_l(x) = \mathrm{E}_x(Y_{1l} - \mu_l), G_{ll'}(x) = \mathrm{E}_x(Y_{1l} - \mu_l)(Y_{1l'} - \mu_{l'})$ and $\delta_{xl}$ is a measurable function from $D$ to $\mathbb{R}$ for each $l = 1, \ldots, \ell$ satisfying the Poisson equation*

$$(I - P)\delta_{xl} = P\Gamma_l(x), \quad \text{(A.8)}$$

*where $I$ denotes the identity kernel and the operators in* (A.8) *are defined according to* (A.2).

**Proof.** By the regeneration method of Markov random walks developed in Ney and Nummelin [23], and following a proof similar to Theorem 17.3.6 in Meyn and Tweedie [22], we have the central limit theorem (A.6). To derive (A.7), we need to show (a) the existence of a finite-valued solution $\delta$ to the Poisson equation (A.8); (b) the uniqueness of $\delta_{yl} - \delta_{xl}$ for all $x, y \in D$ and $l = 1, \ldots, \ell$; and (c) the validity of the variance formula (A.7).

Let $N_\Delta = \inf\{n \geq 0 : X_n = \Delta\}$, and $\delta_{xl} = \mathrm{E}_x(\sum_{k=0}^{N_\Delta} u_l(X_k))$, where $u_l(x) = P\mathrm{E}_x Y_{0l} - \mu_l = \mathrm{E}_x Y_{1l} - \mu_l$. Under the assumption (A.4), $\delta_{xl}$ is well defined, and we have $P\delta_{xl} = \mathrm{E}_x(\sum_{k=1}^{N_\Delta} u_l(X_k))I_{\{x \neq \Delta\}} + \mathrm{E}_\Delta \sum_{k=1}^{T_\Delta} u_l(X_k) = \mathrm{E}_x(\sum_{k=1}^{N_\Delta} u_l(X_k))I_{\{x \neq \Delta\}}$. Therefore, for all $x \in D$ and $l = 1, \ldots, \ell$, $P\delta_{xl} = \mathrm{E}_x(\sum_{k=0}^{N_\Delta} u_l(X_k)) - u_l(x) = \delta_{xl} - P\Gamma_l(x)$, so that the Poisson equation is satisfied, which establishes (a).

The proof of (b) follows from Proposition 17.4.1 of Meyn and Tweedie [22]. We now undertake the proof of (c). By Theorem 17.4.2 of Meyn and Tweedie [22], $\int_D \sum_{k=1}^\infty \Gamma_i(x') \times P^k(x, \mathrm{d}x')$ is finite. Therefore, by a simple generalization of Theorem 3.3 of Billingsley [6], we have

$$\sigma_{ll'}^2 = \mathrm{E}_\pi(Y_{1l} - \mu_l)(Y_{1l'} - \mu_{l'}) + 2\sum_{k=1}^\infty \mathrm{E}_\pi[(Y_{1l} - \mu_l)(Y_{(k+1)l'} - \mu_{l'})]$$

$$= \int_D G_{ll'}(x)\pi(\mathrm{d}x) + \int_D \Gamma_l(x)\tilde{\delta}_{xl'}\pi(\mathrm{d}x) + \int_D \Gamma_l(x')\tilde{\delta}_{x'l'}\pi(\mathrm{d}x'), \quad \text{(A.9)}$$

where

$$\tilde{\delta}_{xl} = \int_D \Gamma_l(x') \sum_{k=1}^{\infty} P^k(x, \mathrm{d}x'). \tag{A.10}$$

Next, we show that $\tilde{\delta}_{xl}$ satisfies the Poisson equation (A.8). That is, for all $x \in D$ and $l = 1, \ldots, \ell$,

$$\begin{aligned}
(I - P)\tilde{\delta}_{xl} &= \int_D \Gamma_l(x') \sum_{k=1}^{\infty} P^k(x, \mathrm{d}x') - \int_D \int_D \Gamma_l(x') P(x, \mathrm{d}z) \sum_{k=1}^{\infty} P^k(z, \mathrm{d}x') \\
&= \int_D \Gamma_l(x') \sum_{k=1}^{\infty} P^k(x, \mathrm{d}x') - \int_D \Gamma_l(x') \sum_{k=1}^{\infty} P^{(k+1)}(x, \mathrm{d}x') \\
&= \int_D \Gamma_l(x') P(x, \mathrm{d}x') = P\Gamma_l(x).
\end{aligned}$$

Write $\delta_l = \delta_{\cdot l}$; then $\delta_l$ and $\Gamma_l$ are measurable functions on $D$. Let $P\delta_l$ be the function and $\pi\Gamma_l\Gamma_{l'}$, $\pi\delta_l$ and $\pi\delta_l\delta_{l'}$ be measures defined according to (A.2), and write $\Gamma_l\delta_l := \Gamma_l(x)\delta_{xl}$. Assuming that $\delta_l$ and $\Gamma_l$ satisfy (A.8), then we have

$$\begin{aligned}
\int_D \int_D &[\Gamma_l(x') - \delta_{xl} + \delta_{x'l}][\Gamma_{l'}(x') - \delta_{xl'} + \delta_{x'l'}]P(x, \mathrm{d}x')\pi(\mathrm{d}x) \\
&= \int_D \int_D (\Gamma_l(x')\Gamma_{l'}(x') - \Gamma_l(x')\delta_{xl'} - \Gamma_{l'}(x')\delta_{xl} + \Gamma_l(x')\delta_{x'l'} + \Gamma_{l'}(x')\delta_{x'l} \\
&\quad + \delta_{xl}\delta_{xl'} + \delta_{x'l}\delta_{x'l'} - \delta_{xl}\delta_{x'l'} - \delta_{x'l}\delta_{xl'})\pi(\mathrm{d}x)P(x, \mathrm{d}x') \\
&= \pi P\Gamma_l\Gamma_{l'} - \pi\delta_{l'}P\Gamma_l - \pi\delta_l P\Gamma_{l'} + \pi P\Gamma_l\delta_{l'} \\
&\quad + \pi P\Gamma_{l'}\delta_l + \pi\delta_l\delta_{l'}P + \pi P\delta_l\delta_{l'} - \pi\delta_l P\delta_{l'} - \pi\delta_{l'}P\delta_l \\
&= \pi\Gamma_l\Gamma_{l'} - \pi\delta_{l'}(I - P)\delta_l - \pi\delta_l(I - P)\delta_{l'} \\
&\quad + \pi\Gamma_l\delta_{l'} + \pi\Gamma_{l'}\delta_l + 2\pi\delta_l\delta_{l'} - \pi\delta_l P\delta_{l'} - \pi\delta_{l'}P\delta_l \\
&= \pi\Gamma_l\Gamma_{l'} + \pi\Gamma_l\delta_{l'} + \pi\Gamma_{l'}\delta_l \\
&= \int_D \Gamma_l(x)\Gamma_{l'}(x)\pi(\mathrm{d}x) + \int_D \Gamma_l(x)\delta_{xl'}\pi(\mathrm{d}x) + \int_D \Gamma_{l'}(x')\delta_{x'l}\pi(\mathrm{d}x'). \tag{A.11}
\end{aligned}$$

By (b), $\delta_{x'l} - \delta_{xl}$ is uniquely determined; then we can replace $\delta_{xl}$ in (A.11) with $\tilde{\delta}_{xl}$ of (A.10). Adding the first term of (A.7) to (A.11) and making use of (A.9) establish the asymptotic variance formula (A.7). $\square$

### A.3. LAN family for Markov random walks

The Markov chain discussed here is assumed to reside on a general state space $D$, whereas the application in (3.4) requires only a finite state space. Let $x \in D$, $\vartheta = (\vartheta_1, \ldots, \vartheta_\ell) \in \mathbb{R}^\ell$, and let $g$ be a bounded measurable function $D$. Define the linear operators $\mathbf{P}_\vartheta$ and $\mathbf{P}$ by

$$(\mathbf{P}_\vartheta g)(x) = \mathrm{E}_x\{\mathrm{e}^{\vartheta \cdot Y_1} g(X_1)\}, \qquad (\mathbf{P}g)(x) = \mathrm{E}_x\{g(X_1)\}, \qquad (\mathrm{A}.12)$$

where "·" denotes the inner product. We assume that $\mathrm{E}_\pi \mathrm{e}^{\vartheta \cdot Y_1} < \infty$ for all $\vartheta \in \Theta \subset \mathbb{R}^\ell$, where $\Theta$ is an open set containing 0.

Under Condition K1, Theorem 4.1 of Ney and Nummelin [23] shows that $\mathbf{P}_\vartheta$ has a simple maximal eigenvalue $\lambda(\vartheta)$ with associated right eigenfunction $r(\cdot; \vartheta)$. Furthermore, there exists a set $F \subset D$ with maximum irreducibility measure $\varphi(F^c) = 0$ such that $\Lambda(\vartheta) = \log \lambda(\vartheta)$ is analytic and strictly convex on $\Theta$, and $r(\cdot; \vartheta)$ is uniformly positive, bounded and analytic on $\Theta$ for each $x \in F$. Now, for $\vartheta \in \Theta$, define the 'twisting' transformation for the transition probability of $\{X_n, n \geq 0\}$,

$$P_\vartheta(x, \mathrm{d}x') = \frac{r(x'; \vartheta)}{r(x; \vartheta)} \mathrm{e}^{-\Lambda(\vartheta) + \vartheta \cdot Y_1} P(x, \mathrm{d}x'). \qquad (\mathrm{A}.13)$$

If the function $\Lambda(\vartheta)$ is normalized so that $\Lambda(0) = \mathrm{d}\Lambda/\mathrm{d}\vartheta|_{\vartheta=0} = 0$, then $P_0 = P$ is the transition probability of the Markov chain $\{X_n, n \geq 0\}$ with invariant probability measure $\pi$.

For all $\vartheta \in \Theta$, let $a(t)$ be a probability distribution on the set of non-negative integers, and let $K_\vartheta^a = \sum_{t=0}^\infty a(t) P_\vartheta^\mathrm{T}$, where $P_\vartheta^\mathrm{T}$ denotes the $t$-step transition of $P_\vartheta$. A set $E \subset D$ is called $\nu_a$-*petite* if there exists a non-trivial measure $\nu_a$ on $\mathcal{D}$ such that $K_\vartheta^a(x, A) \geq \nu_a(A)$ for all $x \in E$ and $A \in \mathcal{D}$.

**Condition K2** *Drift. There exists a function $w: D \to [1, \infty)$, a petite set $E \in \mathcal{D}$, a constant $b < \infty$, and an extended real-valued function $V: D \to [0, \infty]$ such that, for all $x \in D$, $P_\vartheta V(x) \leq V(x) - w(x) + b I_E(x)$, where $P_\vartheta V(x) = \int V(x') P_\vartheta(x, \mathrm{d}x')$ and $I$ denotes the indicator function.*

The following lemma characterizes the constants in (A.13) via a Poisson equation.

**Lemma 2.** *Assume that Conditions K1 and K2 hold for the Markov chain $\{X_n, n \geq 0\}$ with corresponding $V$, $w$ and $b$, such that $\int_D V(x) \pi(\mathrm{d}x) < \infty$. Assume that $\mathrm{E}_\pi \mathrm{e}^{\vartheta \cdot Y_1} < \infty$ for all $\vartheta \in \Theta$. Let $\mu = \mathrm{E}_\pi Y_1$. Then the partial derivatives of $r(\cdot; \vartheta)$ with respect to $\vartheta_k$, $\partial r(x; \vartheta)/\partial \vartheta_k|_{\vartheta=0}$, for $k = 1, \ldots, \ell$, are bounded on $F \subset D$, and are the solutions of the Poisson equation*

$$(I - \mathbf{P})g = \mathbf{P}(E_x Y_1 - \mu),$$

*where $I$ is the identity operator and $\mathbf{P}$ is the operator defined in (A.12).*

The proof of Lemma 2 can be found in Fuh and Hu ([11], Theorem 3).

**Theorem 3.** *Under the assumptions of Lemma 2, let $\vartheta_n = \eta/\sqrt{n}$ and define the transition probability $Q_n^\eta = P_{\vartheta_n}$ through (A.13). Then $Q_n^\eta$ is LAN. In particular,*

$$\lim_{n\to\infty} \frac{\mathrm{d}Q_n^\eta}{\mathrm{d}P} = \exp\left(Z - \frac{1}{2}\eta^\mathrm{T}\Sigma\eta\right),$$

*where $Z$ is a normal random variable with mean zero and variance $\eta^\mathrm{T}\Sigma\eta$.*

**Proof.** By Theorem 4.1 of Ney and Nummelin [23], $\Lambda(\cdot)$ and $r(x,\cdot)$ are analytic on $\Theta$ for each $x \in F \subset D$. A straightforward Taylor expansion gives

$$\Lambda\left(\frac{\eta}{\sqrt{n}}\right) = \Lambda(0) + \frac{\eta}{\sqrt{n}}\Lambda'(0) + \frac{1}{2n}\eta^\mathrm{T}\Lambda''(0)\eta + o\left(\frac{1}{n}\right) = \frac{\eta}{\sqrt{n}}\mu + \frac{1}{2n}\eta^\mathrm{T}\Sigma\eta + o\left(\frac{1}{n}\right),$$

$$\log r\left(x, \frac{\eta}{\sqrt{n}}\right) = \log r(x,0) + \frac{r'(x,0)}{r(x,0)}\frac{\eta}{\sqrt{n}} + o\left(\frac{1}{\sqrt{n}}\right) = r'(x,0)\frac{\eta}{\sqrt{n}} + o\left(\frac{1}{\sqrt{n}}\right).$$

Applying the two preceding expansions to (A.13), we obtain

$$\begin{aligned}\frac{\mathrm{d}Q_n^\eta}{\mathrm{d}P} &= \exp\left\{\frac{\eta}{\sqrt{n}} \cdot \left[S_n - n\Lambda\left(\frac{\eta}{\sqrt{n}}\right)\right] + \log r\left(X_n, \frac{\eta}{\sqrt{n}}\right) - \log r\left(X_0, \frac{\eta}{\sqrt{n}}\right)\right\} \\ &= \exp\left\{\frac{\eta}{\sqrt{n}} \cdot [S_n - n\mu + r'(X_n,0) - r'(X_0,0)] - \frac{1}{2}\eta^\mathrm{T}\Sigma\eta + o_p\left(\frac{1}{\sqrt{n}}\right)\right\}. \quad \text{(A.14)}\end{aligned}$$

The first term in the exponent of (A.14) converges in distribution to a normal random variable with mean zero and variance $\eta^\mathrm{T}\Sigma\eta$. The proof is completed.  $\square$

To apply Theorem 3 to (3.4), we only need to check that $\int f_j(y,\theta)\mathrm{e}^{\theta\cdot y}\nu(\mathrm{d}y) < \infty$ in an open set of $\mathbb{R}^\ell$, as Markov chains in Sections 2 and 3 are of finite state so Conditions K1 and K2 obviously hold.

## Acknowledgements

The first author's research is partially supported by National Science Council of the Republic of China, Grant NSC 95-2118-M-001-008. The second author's research is partially supported by Hong Kong Research Grants Council, Grant HKUST6212/04H. We are grateful to Mr. R.H. Wang for computing assistance.

## References

[1] Albert, P.S. (1991). A two-state Markov mixture model for a time series of epileptic seizure counts. *Biometrics* **47** 1371–1381.
[2] Balish, M., Albert, P.S. and Theodore, W.H. (1991). Seizure frequency in intractable partial epilepsy: A statistical analysis. *Epilepsia* **32** 642–649.

[3] Ball, F. and Rice, J.A. (1992). Stochastic models for ion channels: Introduction and bibliography. *Math. Biosci.* **112** 189–206.

[4] Baum, L.E. and Petrie, T. (1966). Statistical inference for probabilistic functions of finite state Markov chains. *Ann. Math. Statist.* **37** 1554–1563. MR0202264

[5] Bickel, P., Ritov, Y. and Rydén, T. (1998). Asymptotic normality of the maximum likelihood estimator for general hidden Markov models. *Ann. Statist.* **26** 1614–1635. MR1647705

[6] Billingsley, P. (1961). Statistical methods in Markov chains. *Ann. Math. Statist.* **32** 12–40. MR0123420

[7] Cappé, O., Moulines, E. and Rydén, T. (2005). *Inference in Hidden Markov Models.* New York: Springer-Verlag. MR2159833

[8] Davison, A.C. (1988). Discussion of the papers by Hinkley and DiCiccio and Romano. *J. R. Statist. Soc. B* **50** 356–357.

[9] Do, K.A. and Hall, P. (1991). On importance sampling for the bootstrap. *Biometrika* **78** 161–167. MR1118241

[10] Elliott, R., Aggoun, L. and Moore, J. (1995). *Hidden Markov Models: Estimation and Control.* New York: Springer-Verlag. MR1323178

[11] Fuh, C.D. and Hu, I. (2000). Asymptotically efficient strategies for a stochastic scheduling problem with order constraints. *Ann. Statist.* **28** 1670–1695. MR1835036

[12] Fuh, C.D. and Hu, I. (2004). Efficient importance sampling for events of moderate deviations with applications. *Biometrika* **91** 471–490. MR2081314

[13] Ghosh, J.K. (1994). *Higher Order Asymptotics.* NSF-CBMS Regional Conference Series. Hayward, CA: Institute of Mathematical Staistics.

[14] Hall, P. (1992). *The Bootstrap and Edgeworth Expansion.* New York: Springer-Verlag. MR1145237

[15] Johns, M.V. (1988). Importance sampling for bootstrap confidence intervals. *J. Amer. Statist. Assoc.* **83** 709–714. MR0963798

[16] Krogh, A., Brown, M., Mian, I.S., Sjolander, K. and Haussler, D. (1994). Hidden Markov models in computational biology: application to protein modeling. *J. Mol. Biol.* **235** 1501–1531.

[17] Künsch, H.R. (2001). State space and hidden Markov models. In D.E. Barndorff-Nielsen, D.R. Cox and C. Klüppelberg (eds), *Complex Stochastic Systems*, pp. 109–173. Boca Raton: Chapman & Hall/CRC. MR1893412

[18] LeCam, L. and Yang, G.L. (2000). *Asymptotics in Statistics.* New York: Springer-Verlag. MR1784901

[19] Leroux, B.G. (1992). Maximum likelihood estimation for hidden Markov models. *Stochastic Process. Appl.* **40** 127–143. MR1145463

[20] MacDonald, I.L. and Zucchini, W. (1997). *Hidden Markov and Other Models for Discrete-Valued Time Series.* London: Chapman & Hall. MR1692202

[21] Miller, H. (1961). A convexity property in the theory of random variables on a finite Markov chain. *Ann. Math. Statist.* **32** 1260–1270. MR0126886

[22] Meyn, S.P. and Tweedie, R.L. (1993). *Markov Chains and Stochastic Stability.* London: Springer-Verlag. MR1287609

[23] Ney, P. and Nummelin, E. (1987). Markov additive processes I: Eigenvalue properties and limit theorems. *Ann. Probab.* **15** 561–592. MR0885131

[24] Rabiner, L.R. and Juang, B.H. (1993). *Fundamentals of Speech Recognition.* Englewood Cliffs, NJ: Prentice Hall.

[25] Stoffer, D.S. and Wall, D. (1991). Bootstrapping state space models: Gaussian maximum likelihood estimation and the Kalman filter. *J. Amer. Statist. Assoc.* **86** 1024–1033. MR1146350